\documentclass[%
 manuscript=article,
 reprint,
 amsmath,amssymb,
 aps,
 prb,
]{revtex4-1}

\usepackage{chemformula} 
\usepackage[T1]{fontenc} 
\usepackage{graphics}
\usepackage{makecell}
\usepackage{xcolor}
\usepackage{amsmath}
\usepackage{enumerate}
\usepackage[normalem]{ulem}
\usepackage{multirow}
\usepackage[bottom]{footmisc}
\usepackage{bm}
\begin{document}
\newcommand*\mycommand[1]{\texttt{\emph{#1}}}
\newcommand*\rd[1]{{\color{red} #1}}
\newcommand*\bl[1]{{\color{blue} #1}}
\newcommand\mathplus{+}
\newcommand{\snrxn}[0]{S$_\mathrm{N}$2}
\newcommand{\erxn}[0]{E2}
\newcommand{\Ea}[0]{$E_\mathrm{a}$}
\newcommand{\dEa}[0]{$\Delta E_\mathrm{a}$}
\newcommand{\Eas}[0]{$E_\mathrm{a}^{\mathrm{S}}$}
\newcommand{\Eae}[0]{$E_\mathrm{a}^{\mathrm{E}}$}
\newcommand{\deltaml}[0]{$\Delta$\nobreakdash-ML}

\title{Towards the design of chemical reactions: Machine learning barriers of competing mechanisms in reactant space} 

\author{Stefan Heinen}
\author{Guido Falk von Rudorff}
\author{O. Anatole von Lilienfeld}
\email{anatole.vonlilienfeld@univie.ac.at}
\affiliation{University of Vienna, Faculty of Physics, Kolingasse 14-16, AT-1090 Wien, Austria}
\affiliation{Institute of Physical Chemistry and National Center for Computational Design and Discovery of Novel Materials (MARVEL), Department of Chemistry, University of Basel, Klingelbergstrasse 80, CH-4056 Basel, Switzerland}
\begin{abstract}
The interplay of kinetics and thermodynamics governs reactive processes, 
and their control is key in synthesis efforts. 
While sophisticated numerical methods for studying  equilibrium states have well advanced, quantitative predictions of kinetic behavior remain challenging.
We introduce a reactant-to-barrier (R2B) machine learning model that rapidly and accurately infers activation energies and transition state geometries throughout chemical compound space. 
R2B enjoys improving accuracy as training sets grow, and requires as input solely molecular graph information of the reactant. 
We provide numerical evidence for  the applicability of R2B
for two competing text-book reactions relevant to organic synthesis, \erxn\ and \snrxn, 
trained and tested on chemically diverse quantum data from literature.
After training on 1k to 1.8k examples, 
R2B predicts activation energies on average within less than 2.5 kcal/mol with respect to Coupled-Cluster Singles Doubles (CCSD) reference within milliseconds.
Principal component analysis of kernel matrices reveals the hierarchy of the multiple scales underpinning reactivity in chemical space: Nucleophiles and leaving groups, substituents, and pairwise substituent combinations correspond to systematic lowering of eigenvalues. 
Analysis of R2B based predictions of  $\sim$11.5k \erxn\ and \snrxn\ barriers in gas-phase
for previously undocumented reactants indicates that on average \erxn\ is favored in 75\% of all cases, 
and that \snrxn\ becomes likely for nucleophile/leaving group corresponding to chlorine, and for substituents consisting of hydrogen or electron-withdrawing groups.
Experimental reaction design from first principles is enabled thanks to R2B,
which is demonstrated by the construction of decision trees.  
Numerical R2B based results for interatomic distancs and angles of reactant and transition state  geometries suggest that Hammond's postulate is applicable to \snrxn, 
but not to \erxn.
\end{abstract}
\maketitle
%
%
\section{Introduction}
\begin{figure*}[!ht]
        \includegraphics[width=.99\textwidth]{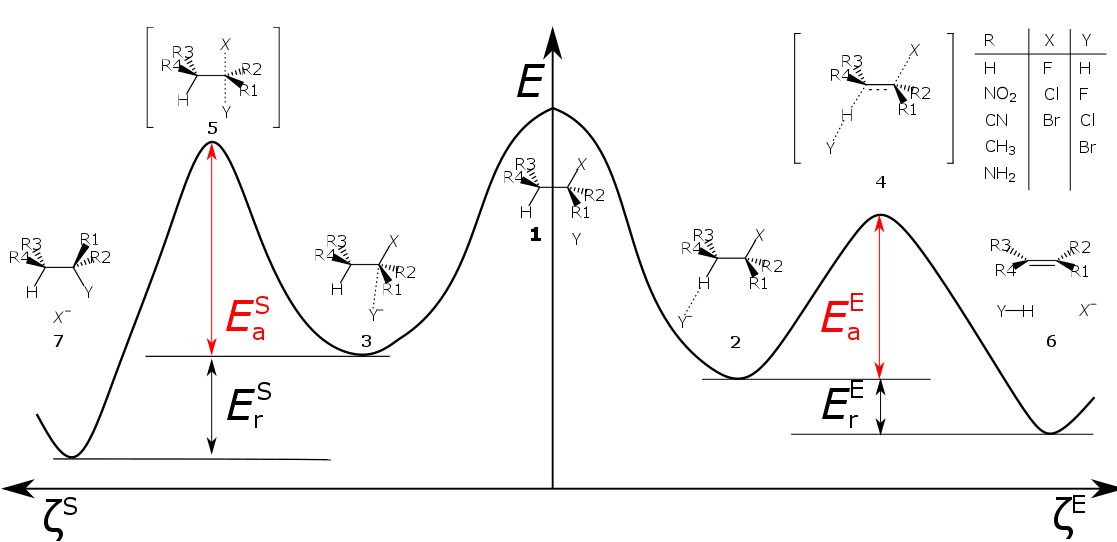}
    \caption{\textbf{Scheme for competing reactions \erxn\ vs. \snrxn.}
    Top row: Transition states \erxn\ (\textbf{4}) and \snrxn\ (\textbf{5}).
    Middle row: Reactant and nucleophile at infinite separation (\textbf{1}). In gas phase the energy of the transition state often lies lower than the energy of the reactants at infinite separation\cite{Vayner_2004}.
    Bottom row: Product geometries at infinite separation (\textbf{6} and \textbf{7}) and reactant complexes (\textbf{2} and \textbf{3}).
    Properties of interest for this work are activation energies \Eae\ and \Eas, reactants, reactant complexes, and transition states.
    Table shows substituents R, leaving groups X, and nucleophiles Y.}
    \label{fig:intro}
\end{figure*}
To accelerate robotic experimental materials synthesis, design, and discovery \cite{cronin2018, guzik2020} a reliable operating system is necessary which can deploy robust virtual models of alternative chemical reaction channels. 
Rapid yet accurate predictions of the kinetic control of reaction outcomes for given reactants and competing reaction channels, however, are still an unsolved problem.
Considerable efforts in quantum chemistry were already directed at the development of automated transition state (TS) searches and chemical reaction paths.
However, calculation of the relevant parts of potential energy surfaces remains a difficult challenge under active research\cite{Young_2020}.
To this end, many TS search algorithms have been introduced which can be grouped into single or double ended methods\cite{Dewyer_2017,Coley_2020}.
An example of the former is the single-ended growing string method\cite{Zimmerman_2015}, which uses only the reactant as starting point and then searches minimum energy paths and transition states.
Double-ended methods such as nudged elastic band \cite{Henkelman_2000, henkelman2002methods} or the two-sided growing string method\cite{Zimmerman2013} employ both reactant and product geometries, to obtain a TS geometry. While successful, both approaches are computationally demanding, and in practice often limited to small systems with mostly single step reactions\cite{grambow2020}.
Recent advances in synthesis planning and modern machine learning techniques 
hold the promise for dramatic acceleration of such numerical 
challenges\cite{bartosz2016, Pflueger2020}.
Already several artificial neural networks to predict reaction outcomes
were introduced (see \cite{MLreact_rev_2020} for a recent review), including work based on molecular orbital interactions of reactive sites\cite{Kayala2011}, molecular fingerprints (template based)\cite{Wei2016}, reaction site identifiers (template free)\cite{NIPS2017_6854, Fooshee2018}, scoring functions in search trees\cite{Segler2018}, sequence to sequence maps\cite{Schwaller2018}, and multiple fingerprint features\cite{Sandfort_2020}.
However, all these machine learning models rely on experimental records, meaning that
they are agnostic of the underlying kinetics 
which are known to be crucial for reliably predicting reaction outcomes.
Neglecting the energetics of chemical reactivity can be problematic, however, due to the reaction rate's exponential dependency on the activation energy (cf. Arrhenius equation).

To use machine learning to go beyond experimental data records and towards  more reliable virtual predictions of reaction outcomes for new chemistries, reaction conditions, catalysts, or solvents, access to substantial and systematic relevant training data of fundamental energetics, e.g.~encoding kinetic or thermodynamic effects, is required\cite{huang2020ab}. 
Very recent first steps in the direction of quantum machine learning applied to reactivity included the prediction of H$_2$ activation barriers of Vaska's complexes~\cite{Friederich2020}, 
the effect of nucleophilic aromatic substitution to reaction barriers~\cite{Jorner2020},
the  temperature dependency of coupled reaction rates\cite{Komp_2020},
or the prediction of enantioselectivity in organocatalysts\cite{enantioClemence2021}.

In this work, we demonstrate how the reactant-to-barrier (R2B) model effectively unifies the two directions (yield vs.~energy) in order to deliver robust predictions of reaction outcomes of competing mechanisms.
We show how R2B can be used to predict and discriminate competing reaction channels among two of the most famous text book reactions in chemistry, \snrxn\ vs.~\erxn \cite{vollhardt} (See Fig.~\ref{fig:intro}) using a quantum data set from the literature encoding thousands of transition states obtained from high-level quantum chemistry\cite{QMrxn20}.
Using our R2B model, we complete the data set for undocumented combinations for which transition state optimizers did not converge. 
We also demonstrate how decision trees based on R2B give actionable suggestions for experiments on how to control which reaction channel dominates, and thus the reaction outcome. 
On the synthetic chemistry side, an analysis of the predicted activation energies, as well as transition state and reactant complex geometries based on our models suggests that Hammond's postulate is not applicable for \erxn.

\section{Methods}
%
%
\subsection{Kernel Ridge Regression}
Ridge regression belongs to the family of supervised learning methods where the input space is mapped to a feature space within which fitting is performed.
The transformation to the feature space is unknown \textit{a priori} and computationally expensive.
To circumvent this problem, the ``kernel trick''\cite{kevinmurphy2012} is applied where the inner product $\langle\mathbf{x}_i, \mathbf{x}_j \rangle$ of the representations of the two compounds $i$ and $j$ are replaced by the so-called kernel function $k(\mathbf{x}_i, \mathbf{x}_j)$.
This results in kernel ridge regression (KRR).
A kernel is a measurement of similarity between two input vectors $\mathbf{x}_i$ and $\mathbf{x}_j$.
In this work, we used the Gaussian kernel:
\begin{equation}
    k(\mathbf{x}_i, \mathbf{x}_j) = \mathrm{exp}\left( -\frac{||\mathbf{x}_i - \mathbf{x}_j||_2^{2}}{2\sigma^{2}}\right) 
\end{equation}
with the length scale hyperparameter $\sigma$ and representation $\mathbf{x}$.
Using the representation of a molecule as input space, KRR learns a mapping function to a property $y^{\rm est}_q(\mathbf{x}_q)$, given a training set of $N$ reference pairs ${(\mathbf{x}_i, y_i)}$.
The property $y^{\rm est}_q(\mathbf{x}_q)$ can be expanded in a kernel-basis set series centered on all the $N$ training instances $i$, 
\begin{equation}
    y^{\rm est}_q(\mathbf{x}_q) = \sum^{N}_i \alpha_i k(\mathbf{x}_i, \mathbf{x}_q)
\end{equation}
where $\{\alpha_i\}$ is the set of regression coefficients which can be obtained as follows:
\begin{equation}
    \bm{\alpha} = (\mathbf{K} + \lambda \mathbf{I})^{-1} \mathbf{y}
\end{equation}
with the regularization strength $\lambda$, the identity matrix $\mathbf{I}$ and the kernel matrix $\mathbf{K}$ with kernel elements $k(\mathbf{x}_i, \mathbf{x}_j)$ for all training compounds.
The kernel ($\mathbf{K}$) within a representation stays the same for both reactions and the difference in the R2B models (${\bm\alpha}$) enters in the change of the label ($\mathbf{y}$) \cite{singleKernel}.
%
%
\subsection{Representations}
Here, we have selected four representations of varying complexity: the Bag of Bonds (BoB)\cite{BobPaper}, spectrum of London\cite{london} and Axilrod-Teller-Muto\cite{AxilrodTeller, Muto} potentials (SLATM), FCHL19\cite{FCHL19} and one-hot encoding\cite{kevinmurphy2012}.

BoB uses the nuclear Coulomb repulsion terms from the Coulomb matrix representation (CM\cite{CM}), and groups them into different bins (so-called bags) for all the different elemental atom pair combinations.
SLATM~\cite{Amons} uses London dispersion contributions as two body term (rather than coulomb repulsion) and Axilrod-Teller-Muto potential as three body term.
While the FCHL18 parameterization accounts for one-body effects in terms of the position of the element in the periodic table (group and period)~\cite{FCHL}, FCHL19 
limits itself to two- and three-body terms
for the sake of computational efficiency~\cite{FCHL19}. 
Its two-body terms contain interatomic distances $R$ scaled by $R^{-4}$, and the three-body terms account for the angular information among all atom triples scaled by $R^{-2}$.

All three geometry-based representations have been tested extensively on close-to-equilibrium structures. 
Since reactive processes, by definition, deal with out of equilibrium structures, we have also included a simple geometry free representation, namely one-hot encoding. 
This representation has also been used to encode amino acids in peptides for artificial neural networks\cite{M_ller_2018, Sp_nig_2019}.
In one-hot encoding, the representation is a vector of zeros and ones (i.e. a bit vector), where only one entry is non zero per feature.
To describe the molecules, we used a bit vector for every substitution site (R$i \in \{1,2,3,4\}$, and one for the nucleophiles (Y) and the leaving group (X), respectively.
This results in a combined vector containing 6 bit vectors of total length of 27 bits.

%
%
\subsection{Training \& Testing: Learning curves}
To train our R2B models, the data set was split into a training set and a test set to optimize the hyperparameters and evaluate the model, respectively.
To get the optimal hyperparameters, we used $k$-fold cross validation\cite{kevinmurphy2012}.
We divide the training data into $k$ folds and for each fold, we trained on all but one fold which was used for evaluating the model.
This procedure was done in an iterative fashion over all the folds.
We then calculated the averaged error over these folds.
This was done for different combinations of hyperparameters $\sigma$ and $\lambda$.

The input for all the geometry based R2B models was the reactants at infinite separation (Figure \ref{fig:intro} compound \textbf{1}).
For each reaction, different reactant conformers (yielding different reactant complexes, Figure \ref{fig:intro} compound \textbf{2} and \textbf{3}) have been reported in the data set~\cite{QMrxn20}. 
To obtain a uniquely defined problem for the ML models, we canonicalized the reactant complexes by always choosing the lowest-lying one from the source data base.
Using compound \textbf{1} the kernel for both reaction channels is the same ($K^{\mathrm{tot}}$), which contains 2 kernels: one for the molecule (\textit{M} and \textit{M'}) and one for the attacking group (\textit{Y} and \textit{Y'}) as shown in equation 4.
Therefore, for both reactions, the same kernel can be used, and the difference in the training enters by the activation energy (\textbf{y}) in equation 3.
\begin{equation}
    K^{\mathrm{tot}} = K(Y, Y') \cdot K(M, M')
\end{equation}
Since one-hot encoding does not depend on the geometry, the kernel can be calculated directly for the entire system.

In order to measure the accuracy of our R2B models, we picked the best set of hyperparameters and trained the model using different training set sizes $N$ and plotted the mean absolute errors (MAE) vs. $N$, resulting in learning curves.
Using learning curves allowed us to see the learning behavior of our R2B models and compare different representations.
The error $\epsilon$ of a consistently improving ML model should decrease linearly for increasing  training set sizes $N$\cite{vapnik}:
\begin{equation}
    \mathrm{log}(\epsilon) = \mathrm{log}(a) - b\cdot \mathrm{log}(N) + \mathrm{HOT}
\end{equation}
where $a$ is the offset (an indicator of how well the selected basis functions fit reality) and $b$ the slope of the learning curve which describes the speed of which the accuracy increases using larger training set sizes. $\mathrm{HOT}$ stands for higher order terms which were neglected in this work, as commonly done.

\subsection{Data \& Scripts}

The data extracted from QMrxn20\cite{QMrxn20} are available on github\cite{gitQMrxn20}.
The scripts used to optimize the hyperparameters and to generate the learning curves are also available in the same git repository.

The data set QMrxn20\cite{QMrxn20} contains 1,286 E2 and 2,361 S$_{\mathrm{N}}2$ machine learned LCCSD activation barriers ($\Delta E_{\mathrm{a}}$).
From these reactions, 529 are overlapping reactions, meaning they start from the same reactant (\textbf{1}) and go over different reactant complexes (\erxn: \textbf{2} and \snrxn: \textbf{3}) towards the corresponding transition states (\erxn: \textbf{4} and \snrxn: \textbf{5}). 
All geometries in the data set had been optimized with MP2/6-311G(d)\cite{Krishnan1980, Curtiss1995, McLean1980, Frisch1984, Clark1983} and subsequently DF-LCCSD/cc-TZVP single point calculations (as implemented in Molpro2018) were performed\cite{Werner2011, Hampel1992, Schuetz2003, Dunning1989, Kendall1992, Wilson1999, Woon1993}.

\section{Results and discussion}
%
%
%
\subsection{Learning Barriers}
\begin{figure}[!ht]
    \includegraphics[width=.49\textwidth]{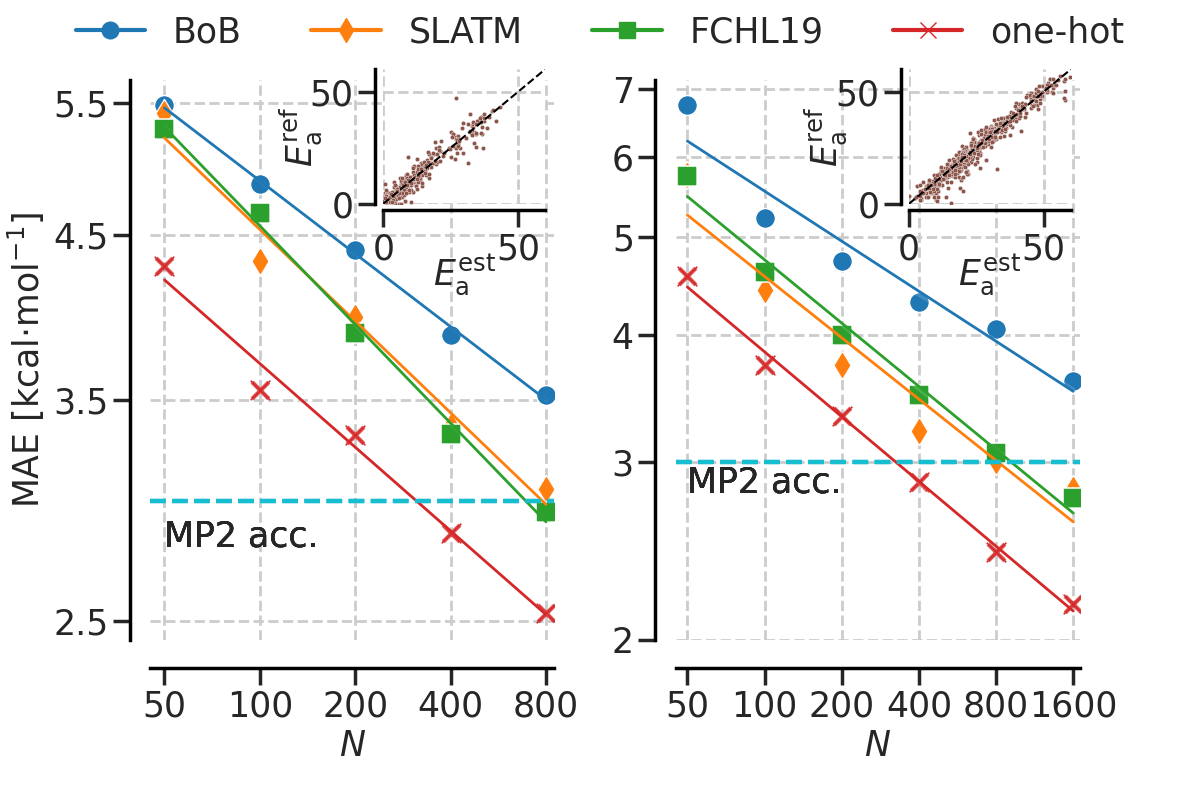}
    \caption{\textbf{Learning curves} Activation energy prediction errors (out-of-sample) as a function of training set size $N$ for activation barriers ($E_{\mathrm{a}}$) of \erxn\ (left) and \snrxn\ (right) using  reactant geometries as inputs only. 
    Results are shown for four representations (BoB, SLATM, FCHL19, one-hot) used within KRR models.  Training data reference level of theory corresponds to DF-LCCSD/cc-pVTZ//MP2/6-311G(d), and estimated MP2 error is denoted as a green dashed horizontal line. 
    Insets: Reference vs. estimated activation barriers using one-hot-based predictions and $R^{2}$ values being 0.89 and 0.94 for \erxn\ (left) and \snrxn\ (right), respectively.}
    \label{fig:lc_lccsd}
\end{figure}

Conventionally, the first principles based prediction of activation energies 
requires the use of sophisticated search-algorithms which iteratively converge towards relevant transition state geometries which satisfy the potential energy saddle-point criterion \cite{Henkelman_2000, henkelman2000improved, Zimmerman2013}. 
The activation energy is then obtained as the energy differences between
reactant and transition state geometry. 
By contrast, our R2B models solely rely on reactant information as input. 
We trained them using aforementioned geometry based representations BoB~\cite{BobPaper}, SLATM~\cite{Amons}, FCHL19~\cite{FCHL19}, as well as one-hot-encoding, to predict activation energies solely based on reactants at infinite separation as input geometries 
(compound \textbf{1} in Figure \ref{fig:intro}). 
Resulting learning curves in Figure~\ref{fig:lc_lccsd} indicate systematically improving activation energy predictions with increasing training set size $N$ for \erxn\ and \snrxn. 
For both mechanisms, the most data-efficient R2B models (one-hot-encoding)
reach prediction errors of 3 kcal/mol with respect to CCSD reference, i.e.~on par
with the deviation of MP2 from CCSD, 
already for less than 300 training instances. 
For 2'000 training instances, the prediction error approaches would 2 kcal/mol. 
Moreover, the lack of convergence suggests that chemical accuracy
(1 kcal/mol) could be reached if several thousand training data points had been available. 
Insets in Figure \ref{fig:lc_lccsd} show true ($E^{\mathrm{ref}}_\mathrm{a}$) vs.~predicted ($E^{\mathrm{est}}_\mathrm{a}$) activation barriers for both reactions. 
Barriers in the range of zero to fifty kcal/mol are predicted with decent correlation coefficients (0.89 and 0.94 for \erxn\ and \snrxn, respectively).  
In short, after training on reference activation energies obtained for
explicit transition state geometries (taken from QMrxn20 data set~\cite{QMrxn20}),
the learning curves in Figure~\ref{fig:lc_lccsd} amount to overwhelming
evidence that it is possible to circumvent the necessity for explicit
transition state structural search when predicting activation energies
for out-of-sample reactants. 

The trends among learning curves in Figure~\ref{fig:lc_lccsd}, are consistent with
literature results for equilibrium structures:
The accuracy improves when going from BoB to SLATM and FCHL19 
for a given training set size~\cite{Anatole2020NatureReview}.
Most surprisingly, however, all R2B models based on geometry dependent representations are less accurate than one-hot encoding. 
While still unique (a necessary requirement for functional R2B models
\cite{fourier_descriptor, Parsaeifard_2020})
one-hot encoding is devoid of any structural information, and its outstanding performance is therefore in direct conflict with the 
commonly made conclusion that a physics inspired functional form of the representation is crucial
for the performance of R2B models~\cite{BAML,Anatole2020NatureReview,rupp_review2021}. 
Relying only on the period and group information in the periodic table to encode composition, other geometry-free representations have also been applied successfully to the study of elpasolite\cite{Faber_2016_elpasolite}, or perovskite\cite{Schmidt_2017_preovskite} crystal structures. 
Here, by contrast, one-hot encoding provides the compositional information for a fixed scaffold.

One can speculate about the reasons for the surprising relative performance of one-hot encoding. 
Due to its inherent lack of resolution which 
prohibits the distinction between reactant and transition state geometry it could be that 
one-hot encoding represents a more efficient basis which effectively maps onto a lower dimensionality with superior learning performance. 
In particular, the inductive effect (practically independent of specific geometric details) 
is known to dominate barrier heights for the types of reactions under consideration\cite{Bragato_2020},
and it is explicitly accounted for through one-hot encoding without imposing the necessity to differentiate it from the configurational degrees of freedom. 

\begin{figure}[!tbp]
      \includegraphics[width=0.49\textwidth]{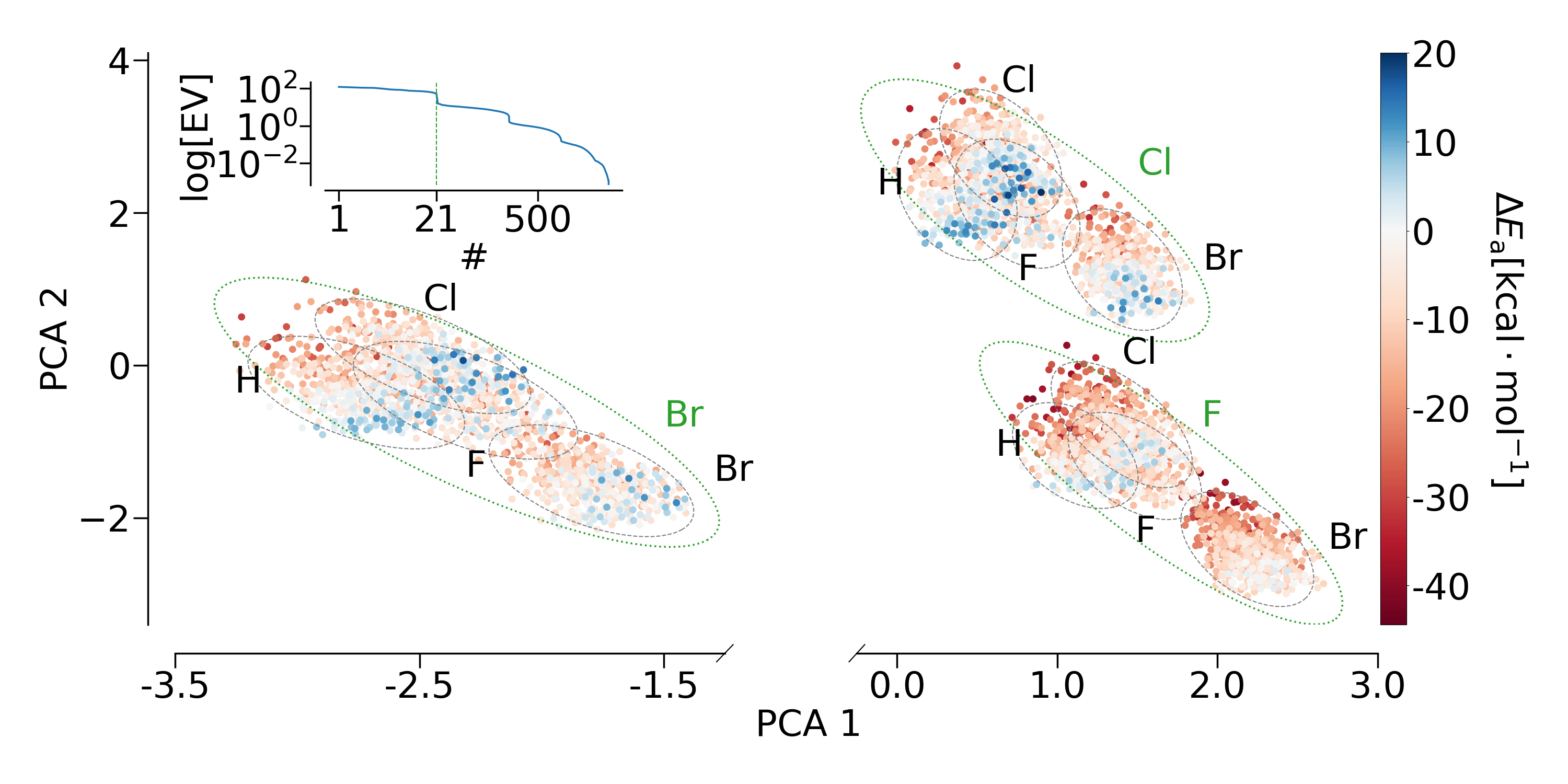}
  \caption{\textbf{Kernel PCA of the training set.}
  Kernel PCA of one hot encoding colored by the energy difference of activation energies of the two reactions $\Delta E_{\mathrm{a}} = E_{\mathrm{a}}^{\mathrm{E}} - E_{\mathrm{a}}^{\mathrm{S}}$.
  Inset: Eigenvalues of the kernel PCA.
  Clusters represent most frequent combinations of leaving groups X (green) and nucleophiles Y (black).
  }
  \label{fig:pca}
\end{figure}

To get an idea of the inner workings of the one-hot encoding model, we performed a principal component analysis (PCA) of the kernel matrix of the predictions which can go either way, i.e.~\erxn\ or \snrxn.
For this subset it is the difference in activation energy which will determine the kinetically stabilized product.
Color coding the first two components by the difference in reference activation barrier labels for the two reactions results in the graphic featured in Fig.~\ref{fig:pca}. 
Confidence ellipsoids of the covariance using Pearson correlation coefficients encode intuitive clusters corresponding to leaving-group/nucleo-phile combinations, and suggest that substituents have less significant effect on trends in activation energies. 
However, the eigenvalue spectrum of the PCA in Figure~\ref{fig:pca} decays rapidly only after the 21 eigenvalue which indicates the number of effective dimensions of the model, and implies that the substituents, alhtough smaller, still have an effect on the activation barrier.
This is consistent with the dimensionality of the one-hot encoding representation: the vector length is 27 (3~X's, 4~Y's and 4$\cdot$5~R's), which is overdetermined,  meaning e.g. the X part of the representation vector consists of three elements F: [1, 0, 0], Cl: [0, 1, 0], or Br: [0, 0, 1].
This could also be uniquely defined with F [0, 0], Cl: [1, 0], Br: [0, 1], which leads a dimension of 21 and is in agreement with the dimensionality of the representation.

\subsection{New barrier estimates }

\begin{figure*}[!ht]
    \includegraphics[width=.99\textwidth]{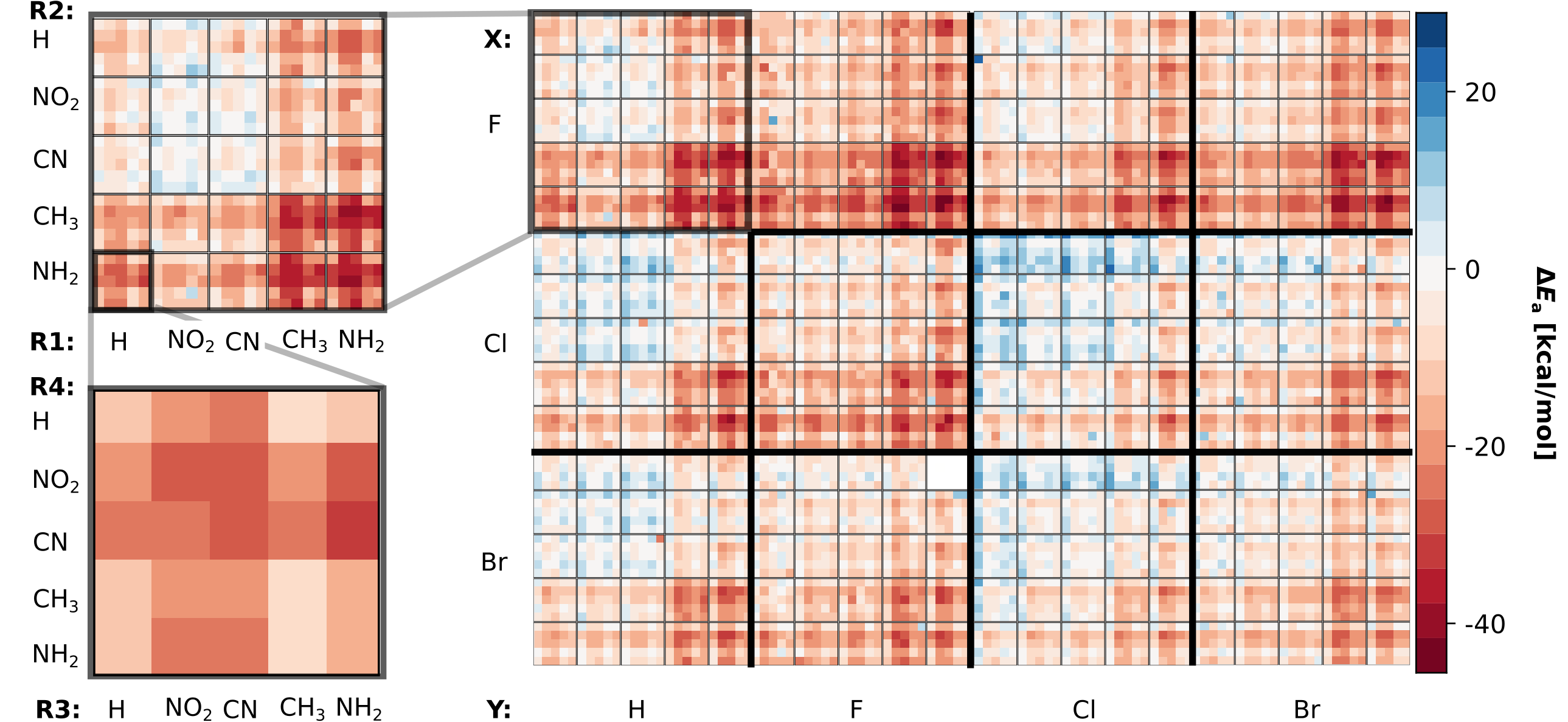}
    \caption{\textbf{Completion of data set using predictions of R2B models} Differences in activation energies ($\Delta E_{\mathrm{a}} = E_{\mathrm{a}}^{\mathrm{E}} - E_{\mathrm{a}}^{\mathrm{S}}$) for all 7,500 reactions (calculated and predicted). Every square stands for a combination of R1-4, X, and Y shown in Figure 1. Positive values denote compounds that undergo a \snrxn\ reaction and negative values lead towards an \erxn\ reaction.}
    \label{fig:heatmap}
\end{figure*}

\begin{figure}[!ht]
    \includegraphics[width=.49\textwidth]{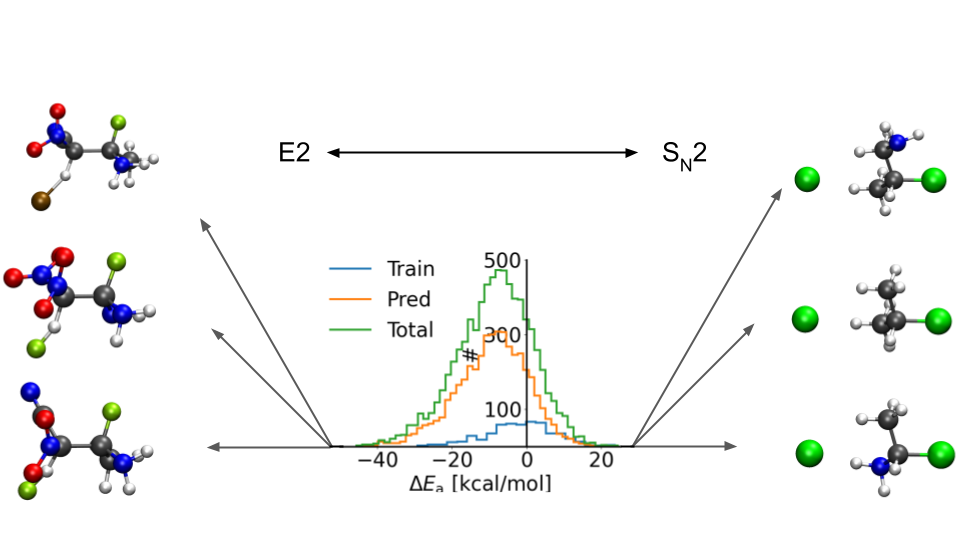}
    \caption{\textbf{Histogram of energy distribution of $\Delta E_{\mathrm{a}}$.} Differences in activation energies ($\Delta E_{\mathrm{a}} = E_{\mathrm{a}}^{\mathrm{E}} - E_{\mathrm{a}}^{\mathrm{S}}$) of 529 overlapping training instances (blue), ~11k predictions (orange) and all 7'500 reactions (green). Molecules of the three highest, respectively lowest barrier differences are shown as molecules.}
    \label{fig:histo}
\end{figure}

Using one-hot encoding (leading to the most performing model) we have trained two models, corresponding to
the 1'286 and 2'361 activation energies of \erxn\ and \snrxn\ transition state geometries, respectively. 
Subsequently, these two models were used to predict 11'353 \erxn\ and \snrxn\ activation barriers for which conventional transition state search methods had failed within the protocol leading up to the training data set~\cite{QMrxn20}.
A summary of the difference in these predicted activation barriers is presented in Figure \ref{fig:heatmap}, where the $x$-axis corresponds to the nucleophiles Y, the $y$-axis to the leaving groups X.
For every combination of X and Y, there are 5$\cdot$5 squares for the functional groups at position R1 and R2.
Within these, there are again 5$\cdot$5 squares belonging to R3 and R4.
Each of the squares represents one reaction for a given combination of R1-4, X, and Y.
Simple heuristic reactivity rules emerge from inspection of these results: If the nucleophile and the leaving group are Cl, the preferred reaction is \snrxn. If the nucleophile and the leaving group are F, the preferred reaction is \erxn. 
The functional groups at positions R1 and R2 favour the \erxn\ due to their electron donating properties which disfavour a nucleophilic back side attack in the \snrxn\ reaction.
A comprehensive overview is shown in Fig.~\ref{fig:heatmap}. 
The same rules can be observed in Figure \ref{fig:histo} which shows the distribution of the differences in activation barrier ($\Delta E^{\mathrm{a}}$) of the training, predicted and total data set.
The molecules of the extreme cases, largest difference in activation energies, are shown for both reactions, \erxn\ (left) and \snrxn\ (right). 
Figure \ref{fig:histo} shows a favourization of the \erxn\ reaction of a rate of roughly 75\%.
These results have to be taken with caution, since this shift in \erxn\ can also have occurred due to the composition of the molecules in the training set, as well as the choice of small functional groups that minimizes steric effects.
A more detailed discussion of the training, the data set completion with the R2B model, and trends can be found in the SI.

\subsection{Design rule extraction}
\begin{figure*}[!ht]
    \includegraphics[width=.89\textwidth]{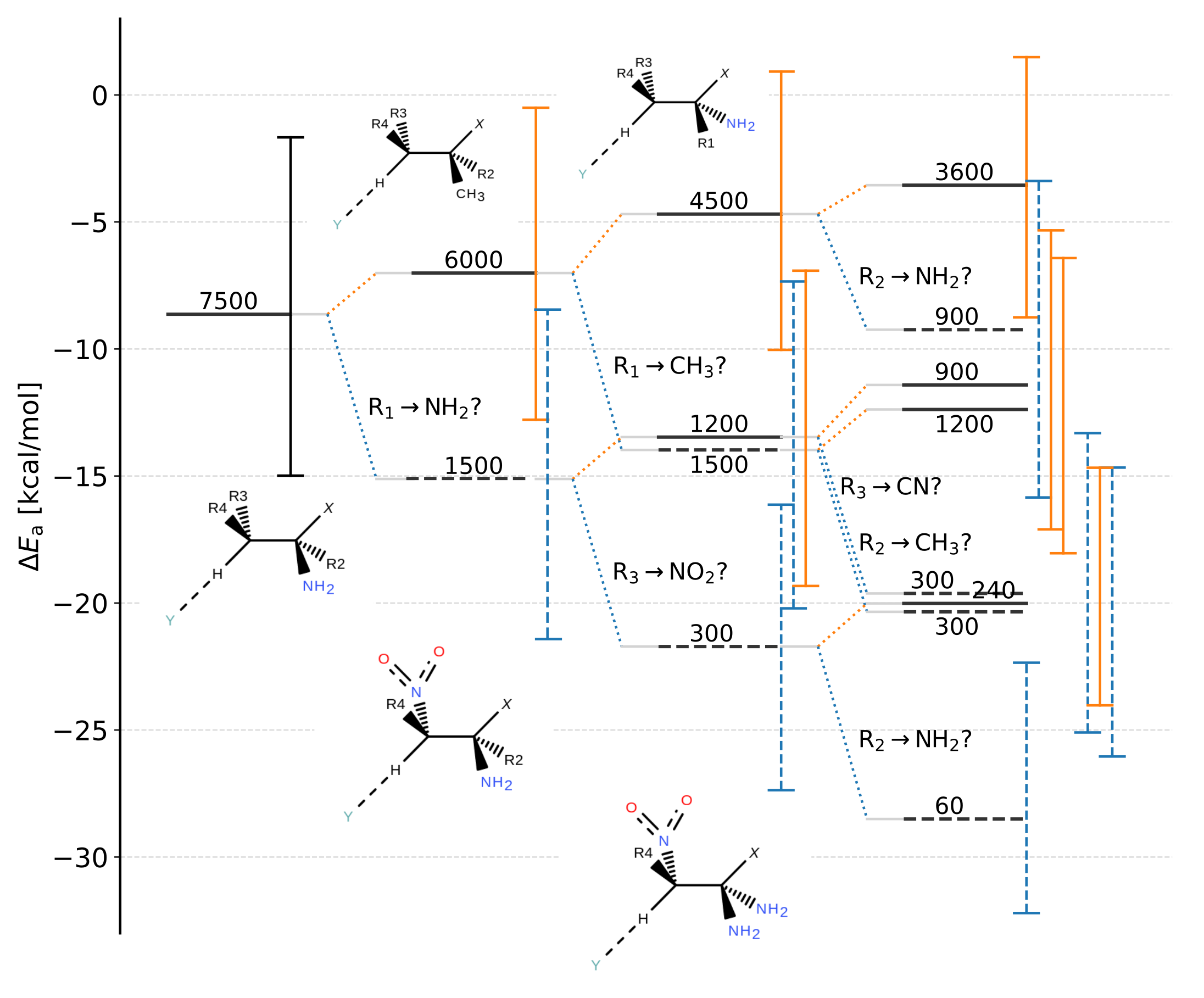}
    \caption{\textbf{Decision tree using extracted rules and design guidelines.}
   Decision tree using the R2B estimated activation barriers to predict changes in barrier heights by starting at all reactions (first energy level on the left) and subsequently apply changes by substituting functional groups, leaving groups and nucleophiles with E2 as an example.
   Blue dotted lines refer to an accepted change meaning only compounds containing this substituents at the position are considered.
   Orange dotted lines refer to substitution declined, meaning all compounds except the decision are kept.
   Vertical lines on the right of energy levels denote the minimum first (lower limit), and the third (upper limit) quartile of a box plot over the energy range.
   Numbers above energy levels correspond to the number of compounds left after the decision.
   Lewis structures resemble the decision in question.
    }
    \label{fig:dtree}
\end{figure*}

So far, most studies based on artificial neural networks aimed at predicting chemical reactions using experimental data do not account for the kinetics of reactions. 
It is well known, however, that activation barriers are crucial for chemical synthesis and retrosynthesis planning.
This is exemplified by a decision tree for the competing reactions \erxn\ and \snrxn\ in Figure \ref{fig:dtree}.
The goal of such trees is to improve the search for better reaction pathways (lower activation barriers), by showing the estimated change in energy when changing functional groups, leaving groups, or nucleophiles.
To extract such rules for the design problem, a large and consistent reaction data set is needed.
After completing the data set\cite{QMrxn20}, we are now able to identify (given a desired product) the estimated changes in the activation barrier, when subtituting specific functional groups, leaving groups, or nucleophiles.
This way, the yield of chemical reactions can be optimized by getting insights of the effects that functional groups have on a certain molecule.
Furthermore, this insight could be used to direct reactions towards the desired product.
Figure \ref{fig:dtree} shows such a possible decision tree to determine the change in barriers while exchanging substituents.
Starting from the total data set (left energy level), the first decision considers the functional group NH$_2$ at position R1.
Going down the tree means accepting the suggested change and the respective compounds, while going up means declining and removing these compounds from the data.
Depending on which product is sought after, hints to improve the energy path can be found while constantly accepting (going down) or declining (going up) the tree.
For example, if the desired reaction is \erxn, then the best way is to go down on the tree (decision accepted) which adds electron withdrawing groups to the R3 and R4 position, as well as electron donating groups to R1 and R2.
In Figure \ref{fig:dtree} the first decision redirects the barrier towards \erxn\ about $\sim$8 kcal/mol by adding an electron withdrawing group (NO$_2$) on the $\alpha$-carbon.
On the other hand, electron donating group at the $\beta$-carbon favour the \erxn\ reaction because they facilitate the abstraction of the leaving group, which is shown in the second and the third decision, where NH$_2$ was added in both positions, R1 and R2. 
In addition to the R2B predictions, which tell you the outcome of a specific combination of one reaction, a decision trees gives simple rules as an coarsened aggregation that can be used in reaction design to achieve a desired outcome.
%
%
\subsection{Estimates of reactant and transition state geometries}
\label{chap:tsgeomlearn}

\begin{figure*}[!tbp]
      \includegraphics[width=0.99\textwidth]{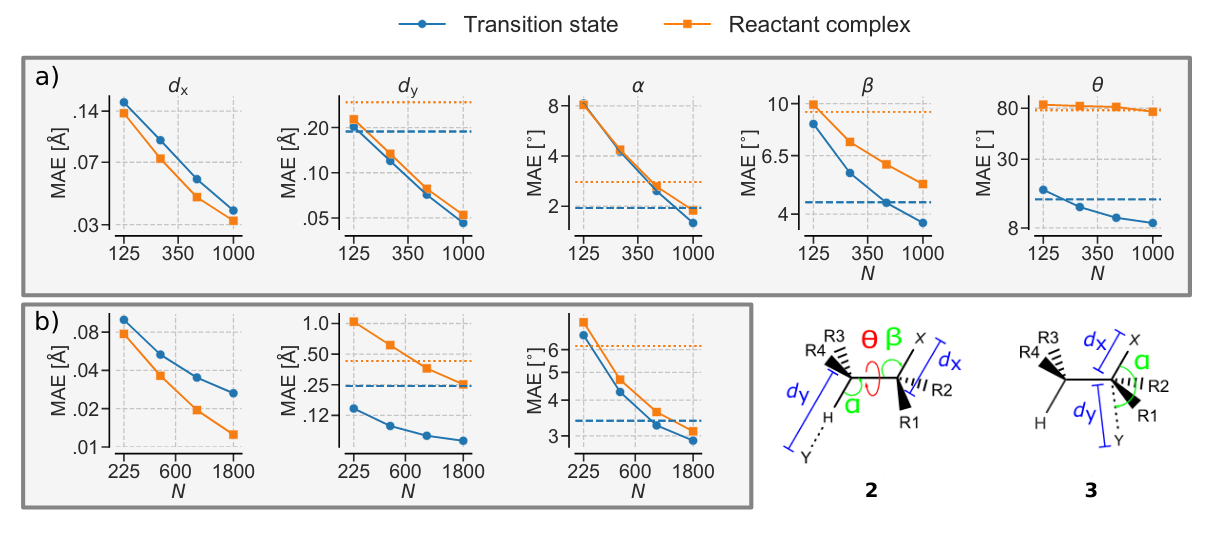}
  \caption{\textbf{Model evaluation of geometrical properties using learning curves.} Test errors (MAE) of distances $d_{\mathrm{x,y}}$, angles $\alpha$ and $\beta$ and dihedrals $\theta$ for \erxn\ (top row) and \snrxn\ (bottom row).
  Horizontal lines correspond to the null model.
  Molecules bottom row show learned properties of the \erxn\ reaction (\textbf{2}) and the \snrxn\ reaction (\textbf{3}) for both structures, reactant complex and transition state.}
  \label{fig:lc_geom}
\end{figure*}

Additionally to barriers, we analysed the geometries of the transition states as well as the geometries of the reactant complexes \cite{QMrxn20}.
Choosing key geometrical parameters, such as distances, angles, and dihedrals, we were able to train R2B models to learn these properties using the one-hot encoding as representation.
These parameters were extracted from the ethylene scaffold defining the key positions of the substituents, leaving groups, and nucleophiles shown in Figure \ref{fig:lc_geom} compounds \textbf{2} and \textbf{3} for the \erxn\ and \snrxn\ reaction, respectively.

The parameters for the \erxn\ reaction are the C-X distance $d_\mathrm{x}$, the C-Y distance $d_\mathrm{y}$, the X-C-C angle $\alpha$, the C-C-Y angle $\beta$, and the X-C-C-Y dihedral $\theta$.
Similarly for \snrxn, we have the C-X distance $d_\mathrm{x}$, the C-Y distance $d_\mathrm{y}$, and the X-C-Y angle $\alpha$.
For every parameter, a separate model was trained using the one-hot representation.
Although this representation does not contain any geometrical information, learning was achieved for every parameter.
Figure \ref{fig:lc_geom} shows the learning curves and as horizontal dashed lines the null model which uses the mean of the training set for predictions.
In the same way as for the transition state geometries, we also trained a model for the reactant complexes.
Figure \ref{fig:lc_geom} shows the learning curves for both, transition states and reactant complexes.
The results for both geometries are similar except for the dihedral of the reactant complexes.
The poor performance results from the conformer search of the reactants.
Compared to bond distances, dihedrals have multiple local minima which  leads to larger differences between the reactant and transition state structures.
The variance of the dihedrals are significantly higher which makes the learning task much harder.
The one-hot representation does not contain any geometrical information and therefore is not able to learn the different geometries only using information about the constitution (R's, X's, and Y's) of the reactant complexes.

%
%
\subsection{Hammond's postulate}
\begin{figure*}[!tb]
      \includegraphics[width=0.99\textwidth]{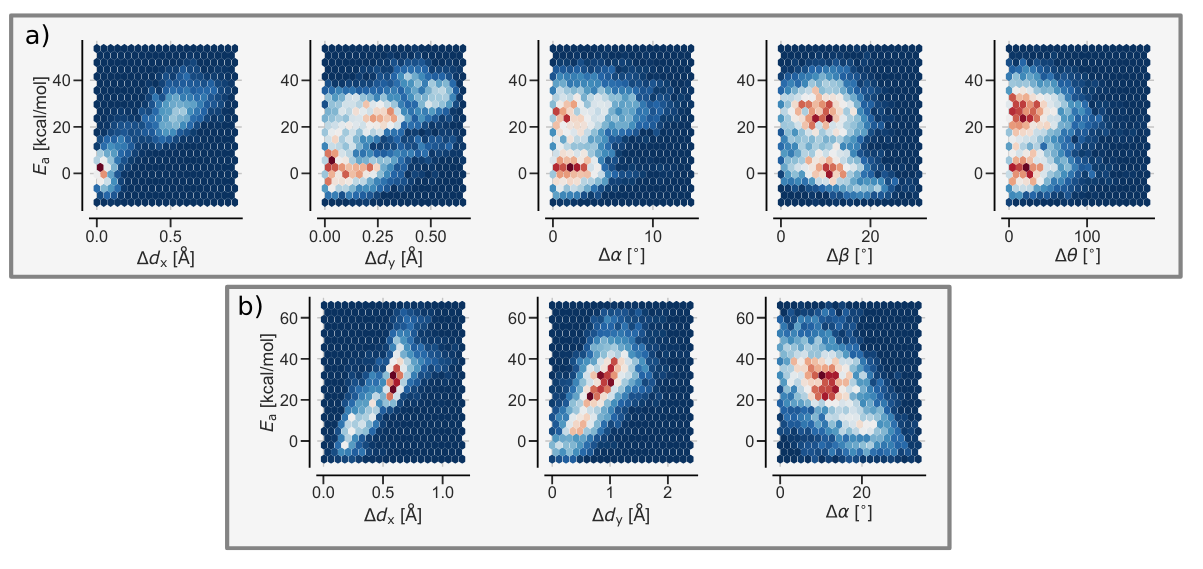}
  \caption{\textbf{Illustration of Hammond's postulate.}
  Comparing structural differences (distances, angles, and dihedrals) to activation energies $E_{\mathrm{a}}$ using five \erxn\ (top row) and three \snrxn\ (bottom row) properties.
  Linear behaviour confirming Hammond's postulate is observed for distances ($d_{\mathrm{x, y}}$) in the \snrxn\ reaction.
  Lack of linearity for the \erxn\ reaction reveals limits of Hammond's postulate when applied to reactant complex conformers.
  Properties learned are displayed in Figure \ref{fig:lc_geom}.}
  \label{fig:hexbin}
\end{figure*}

To investigate Hammond's postulate we took the difference in the predicted geometries for all 7,500 reactions for the five and three parameters for the \erxn\ and the \snrxn\ reaction, respectively.
Then we plotted these values against the activation energies of both reactions \Eae\ and \Eas\ (Figure \ref{fig:hexbin}).
The distances $\Delta d_{\mathrm{x}}$ correlate well with the energies.
This is explained by the leaving group that is bonded to the carbon atom in the reactant complex and only small changes in distance happens moving towards the transition state geometry. 
For the \snrxn\ reaction, the backside attack of the nucleophile does not allow a broad distribution of angles in the reactant complex and the transition state.
Moreover, the changes in geometry between the reactant complex and the transition state are modest.
Therefore, the parameter $\Delta d_{\mathrm{y}}$ for the \snrxn\ correlates well with the activation energy \Eas.
The attack of the nucleophile on the hydrogen atom (\erxn\ reaction) allows for a much broader distribution of the position of the nucleophile in the transition state.
This makes the learning problem more difficult, especially for a representation not including geometrical information.
Therefore predictions for the dihedrals contain large errors.

Angles and dihedrals correlate very poorly with the activation energies because of the low barriers between the different minima along a dihedral for a molecule.
This leads to larger differences in these parameters comparing reactant complexes and transition state geometries.

Hammond's postulate typically holds for the end points of an intrinsic reaction coordinate (IRC) calculation\cite{PeirGarca2003,Zhang2004,Shiroudi2015} which leads to a local minimum close to the transition state.
Therefore, the reactant only needs a few reorganisations towards the transition state.
For geometries that are farther away from the transition state (such as in our \erxn), Hammond's postulate cannot hold anymore.
This means that even though more reorganization steps towards a transition state have to be made, the activation energy is not affected anymore. As a consequence, Hammond's postulate is no longer applicable. 

%
\section{Conclusion}
We have introduced a new machine learning model dubbed Reactant-To-Barrier (R2B) to predict activation barriers using reactants as input only. 
This approach renders the model practically useful, as the dependency on the transition state geometry is only implicitly obtained at the training stage, and not explicitly required for querying the model. 
We find that one-hot-encoding, the trivial geometry free based representation, yields even better results than geometry based representations designed for equilibrium structures. 
As such, our results indicate that accounting only for the combinations of functional groups, leaving group, and nucleophile of the reaction is sufficient for promising data-efficiency of the model.
Using R2B predictions, we completed the reaction space of QMrxn20\cite{QMrxn20}.
Future work could include delta ML\cite{deltaML2015} to improve these results even further, as corroborated by preliminary results in Ref.~\cite{QMrxn20}, 
further improvements on the representation (as recently found to lead to improved barrier predictions for enantioselectivity in metal-organic catalysts~\cite{enantioClemence2021}),
or the inclusion of catalytic or solvent effects~\cite{fml}. 

Using R2B predicted activation barriers, we have also introduced the notion of a decision tree, enabling the design and discrimination of either reaction channel encoded in the data.
Such trees systematically extract the information hidden in the data and the model regarding the combinatorial many-body effects of functional groups, leaving groups, and nucleophiles which result in one chemical reaction being favoured over the other. As such, they enable the control of chemical reactions in the design space spanned by reactants. 
Finally, we also report on geometries of the reactant complexes consisting of different conformers, as well as on R2B based transition state geometry predictions.
Using these results, we discuss the limitations of Hammond's postulate which does not hold for the E2 reactant complexes stored in the QMrxn20 data set\cite{QMrxn20}.

\section*{Acknowledgement}
This project has received funding from the European Union's Horizon 2020 research and
innovation program under Grant Agreements \#952165
and \#957189.
This project has received funding from the European Research Council (ERC) under the European Union’s Horizon 2020 research and innovation program (grant agreement No. 772834).
This result only reflects the
author's view and the EU is not responsible for any use that may be made of the
information it contains.
This work was partly supported by the NCCR MARVEL, funded by the Swiss National Science Foundation.

\bibliography{literature}
\end{document}